\documentclass[letterpaper,11pt]{article}
\pdfoutput=1
\usepackage{xcolor}
\usepackage{jheppub}
\usepackage{mathtools}
\usepackage{braket}
\usepackage{appendix}
\newcommand\numberthis{\addtocounter{equation}{1}\tag{\theequation}}
\usepackage[linewidth=1pt]{mdframed}
\usepackage{cleveref}

\newcommand{\DD}{\mathrm{d}}
\renewcommand{\.}{\cdot}
\newcommand{\defined}{\coloneqq}
\newcommand{\e}{\epsilon}
\renewcommand{\d}{\partial}
\newcommand{\at}[1]{\bigg\vert_{#1}}

\renewcommand{\O}{\mathcal{O}}
\renewcommand{\d}{\partial}

\title{Angular momentum of the asymptotic electromagnetic field in the classical scattering of charged particles}

\author[a]{Rishabh Bhardwaj,}
\author[a]{Luke Lippstreu}

\affiliation[a]{Department of Physics,
	Brown University,
	Providence, RI 02912, USA}
\emailAdd{rishabh\_bhardwaj@brown.edu}
\emailAdd{luke\_lippstreu@brown.edu}

\abstract{We compute the angular momentum of the electromagnetic field on a late time Cauchy surface with an arbitrary constant normal vector relevant for the classical scattering of charged particles. We find a time independent contribution to the angular momentum. This demonstrates that every charged particle scattering event is accompanied by a net shift in the angular momentum of the electromagnetic field. We speculate that this shift is related to a subleading electromagnetic memory effect. We argue that this asymptotic angular momentum should be included in the description of the asymptotic states in quantum theories containing infrared divergences. We demonstrate that the Lorentz covariance of the asymptotic electromagnetic angular momentum can only be exhibited upon making reference to the Cauchy slice's normal vector. }

\begin{document}
\maketitle
\section{Introduction}
Quantum theories containing massless particles in spacetime dimensions four or less exhibit infrared (IR) divergences due to the asymptotic states failing to converge appropriately towards free particle states. It has been known since the late 1960's that the asymptotic states instead approach what are often referred to as dressed Faddeev-Kulish (FK) states \cite{dollard,Chung:1965zza,Kibble:1968sfb,Kulish:1970ut}. The exact details of these states remain poorly understood as evidenced by the lack of explicit IR finite amplitudes calculated using these states. Progress on theoretical questions of great importance have often been thwarted due to the lack of an unambiguous framework for the calculation of infrared finite amplitudes\footnote{Note that the treatment of IR divergences by summing over physically indistinguishable processes is formulated on the level of cross-sections not amplitudes.}, see e.g. \cite{Caron-Huot:2022jli,Caron-Huot:2022ugt} where the authors obtained causality and unitarity constraints on Wilson EFT coefficients for gravity up to the ambiguity of infrared logarithms.  In order to study questions of causality and unitarity in four dimensions it would be desirable to further clarify the properties of the asymptotic states so that a less ambiguous and more efficient approach to the calculation of IR finite amplitudes can be developed. As a step in this direction, in this paper we provide an answer to the following question:
\begin{center}
    \it{What are the correct quantum numbers to assign to the asymptotic states in theories containing infrared divergences?}
\end{center}
We do so by calculating the classical angular momentum of the electromagnetic (EM) field on a late time Cauchy slice in the presence of a charged particle scattering event. We find, see (\ref{mainresult}-\ref{secondmainresult}) for the main result, that the EM angular momentum asymptotes to a non-zero, time independent value. Although the computation is classical, the result suggests the following answer to the above posed question:  charged particle momentum eigenstates $\ket{\set{p_i^{\mu},\sigma_i,e_i}}$ should be appended by states which account for this asymptotic angular momentum in the EM field. Indeed, this conclusion is in line with the recent work \cite{DiVecchia:2022owy} where it was shown that the FK states carry a non-zero angular momentum. We hope that this insight may lead to future work which alleviates both the infinite ambiguity in the choice of dressings for the FK states, as well as the presence of an infrared cut-off in the FK states which obscures Poincar\'e covariance. Our work may also be of interest to communities studying the infrared triangle of asymptotic symmetries, soft theorems and memory effects. Both Low's subleading soft photon theorem \cite{Low:1954kd,Low:1958sn,Burnett:1967km,Gell-Mann:1954wra,Lysov:2014csa}, as well as the one-loop corrections to the soft photon theorem \cite{Sahoo:2018lxl}, involve terms which couple the soft photon to the angular momentum of the charged particles. As our results indicate that angular momentum conservation requires there to be a net shift in the low energy modes of the angular momentum tensor of the EM field accompanying any charged particle scattering event, it would be of interest to investigate whether this shift is related to loop corrections to the electromagnetic memory effect \cite{Bieri:2013hqa,Pasterski:2015zua,AtulBhatkar:2020hqz} or the memory effect associated to Low's subleading soft theorem \cite{Hirai:2018ijc, Campiglia:2016hvg,Himwich,Campiglia:2019wxe,Donnay:2022sdg}.  \newline
Our work is a follow-up on \cite{Gralla:2021eoi} where the authors calculated the late time angular momentum of the EM field in the presence of charged particles.   The calculation presented in \cite{Gralla:2021eoi} was done in the center of momentum (COM) frame on an equal time Cauchy slice. In this paper we present the covariant form of their results by working in an arbitrary frame using a Cauchy slice with a constant normal vector. By doing the computation on an arbitrary Cauchy slice, our results highlight the interesting feature that the EM angular momentum depends on the details of the Cauchy surface. Our results are in agreement with those in \cite{Gralla:2021eoi} when evaluated in the COM frame on an equal time Cauchy slice. \newline
Our work was motivated by Zwanziger's description of dyons \cite{Zwanziger:1972sx}, see also the recent follow-ups \cite{Csaki1,Csaki2,Lippstreu:2021avq}. It is well known that in the presence of dyons the electromagnetic field contains a time independent angular momentum at asymptotic times. Zwanziger used this fact to modify the Poincar\'e transformation properties of dyons from that of free particles', and in doing so was able to calculate the scattering amplitudes of dyons whilst circumventing many of the ambiguities that otherwise plague their calculation. In a form made precise below our main result (\ref{mainresult}-\ref{secondmainresult}), the angular momentum in the EM field due to charged particles is essentially boost-like in nature, whereas the angular momentum of dyons is essentially rotational in nature. It would be of interest to repeat Zwanziger's analysis for the boost-like angular momentum found herein.\newline
The outline of the paper is as follows: in section \ref{sect:setup} we describe our setup of classical charged particle scattering. In section \ref{sect:emangularmomentum} we present our main result, the angular momentum of the asymptotic electromagnetic field, and discuss some of its qualitative features. In section \ref{sect:mech} we compute the mechanical angular momentum of the charged particles to demonstrate that our results are consistent with total angular momentum conservation. Appendices our dedicated to the evaluation of some of the relevant integrals.\newline 
\textbf{Conventions:} We work in $(+---)$ signature throughout. Our anti-symmetrization bracket $a^{[\mu}b^{\nu]}=a^{\mu}b^{\nu}-a^{\nu}b^{\mu}$ does not contain a factor of $\frac{1}{2}$.
\section{Setup}\label{sect:setup}
Consider an outgoing asymptotic state made up of particles carrying charges $e_i$, each in a wave packet concentrated around a momentum $p_i^{\mu}$ . At sufficiently late times we expect that such states are accurately modelled by classical currents
\begin{equation}
    J^{\mu}_{i}=e_iu^{\mu}_{i}\int_0^{\infty}\DD \tau \,\,\delta^{4}\Big(x^{\mu}-u^{\mu}_{i}\tau-b_i^{\mu}+\O(\log\tau)\Big),\qquad u^{\mu}_{i}=\frac{p^{\mu}_{i}}{m_{i}}\label{currents},
\end{equation}
where $u_i^{\mu}$ is the four velocity $u_i^2=1$ of particle $i$, $b_i^{\mu}$ is it's impact parameter, and the bounds of integration are appropriate for an out state. The trajectories receive order $\ln t$ corrections to linear motion due their late interaction with other particles which we account for later. Using the retarded Green's function\footnote{Here we deviate from the quantum theory by not using the Feynman propagator.} the electromagnetic field produced by particle $i$ reads
\begin{equation}
  \lim_{t\rightarrow \infty}  F^{\mu\nu}_{i}=\frac{e_i}{4\pi}\frac{x_{}^{{[}\mu}{u_i}^{\nu{]}}}{\Big((u_i\. x)^2-x^2\Big)^{\frac{3}{2}}}\theta(x^2)+\O(t^{-3}\log(t)),\qquad x^{{[}\mu}{u_i}^{\nu{]}}\defined x^{\mu}u^{\nu}_i-x^{\nu}u^{\mu}_i\label{EMfieldofpointcharge}.
\end{equation}
The Heaviside $\theta(x^2)$ is due to our use of the retarded propagator. Note that the field strength (\ref{EMfieldofpointcharge}) vanishes as $t^{-2}$ at late times as expected from dimensional considerations. The impact parameter $b_i^{\mu}$ is an order $\O(t^{-3})$ correction to (\ref{EMfieldofpointcharge}) and consequently does not affect the EM angular momentum at late times. Similarly the $\log t$ corrections to the particle trajectories provide an order $\O(t^{-3}\log(t))$ correction to (\ref{EMfieldofpointcharge}) and hence do not affect the late time angular momentum of the EM field, but does affect the mechanical angular momentum.  
The symmetrized energy-momentum and angular momentum densities of the EM field respectively are
\begin{gather}
   T^{\alpha\beta}_{\text{E.M.}}=\Big(g^{\alpha\mu}F_{\mu\lambda}F^{\lambda \beta}+\frac{1}{4}g^{\alpha\beta}F_{\mu\lambda}F^{\mu\lambda}\Big)\label{energymomentumtensor}\\
   M^{\alpha\beta\gamma}_{\text{E.M.}}=T_{\text{E.M.}}^{\alpha[\beta}x^{\gamma]}.
\end{gather}
For the fields (\ref{EMfieldofpointcharge}) these scale at late times respectively as $t^{-4}$ and $t^{-3}$.
We will integrate these densities on a spacelike Cauchy hypersurface $\Sigma_n$ specified by the conditions
\begin{equation}
    n\.x=c,\qquad n^2=1,\qquad n^{0}>0,\qquad c\gg 1 \label{cauchy surface}
\end{equation}
where $n^{\mu}$ is the spacetime independent normal vector to the Cauchy surface, and $c$ is a spacetime independent scalar which we take to be large. For example, choosing $n^{\mu}=(1,\vec{0})$ would correspond to an equal time Cauchy slice, and $c$ would then then be the time of the Cauchy slice. When one computes the flux of a conserved quantity through a Cauchy slice, the result is independent of the hypersurface, hence one can usually work with the most convenient choice of hypersurface without loss of generality . However one of the results of this paper will be to show that the Lorentz covariance of the late time angular momentum of the EM field can only be exhibited if one works on the arbitrary linear Cauchy slice (\ref{cauchy surface}). The EM energy-momentum and angular momentum on such a hypersurface respectively read
\begin{gather}
    P^{\beta}_{\text{E.M.}}=\int_{\Sigma_n}\DD^3\sigma \sqrt{|g_{\sigma\sigma}|}\,\,n_{\alpha}T^{\alpha\beta}\\
    J^{\beta\gamma}_{\text{E.M.}}=\int_{\Sigma_n}\DD^3\sigma \sqrt{|g_{\sigma\sigma}|}\,\,n_{\alpha}M^{\alpha\beta\gamma}\label{angular momentum}
\end{gather}
where $\sigma_i$ denote co-ordinates adapted to the Cauchy surface and $\sqrt{|g_{\sigma\sigma}|}$ is the induced measure. Due to the $\theta(x^2)$ in (\ref{EMfieldofpointcharge}) we expect the integrals to occur over a volume of order $t^3$. We therefore expect the electromagnetic momentum to vanish as $P^{\alpha}_{\text{E.M.}}\sim t^{-1}$ at late times.  The angular momentum contains an extra power of $x^{\mu}$, we therefore expect it will contain a finite contribution $J^{\beta\gamma}_{\text{E.M.}}\sim t^{0}$. For our Cauchy surface (\ref{cauchy surface}) all power counting arguments carry over where one instead power counts in terms of the large parameter $c$.\newline
Each field (\ref{EMfieldofpointcharge}) exhibits the familiar $\frac{1}{r^2}$ UV divergence as $x^{\mu}$ approaches the location of the source particle. Consequently the order $e_i^2$ terms in the angular momentum tensor (\ref{angular momentum}) will diverge logarithmically whereas the order $e_ie_j$ terms for $i\neq j$ are finite. Such UV divergences are attributable to self-energy corrections which are treated by mass renormalization. As we are not attempting to calculate self-energy corrections we will drop these self-energy contributions to the angular momentum tensor. A further justification for this is that the conservation laws of angular momentum will hold independently for the order $e_i^2$ and $e_ie_j$ terms. 
\section{Electromagnetic angular momentum in the asymptotic states}\label{sect:emangularmomentum}
Using the field strengths (\ref{EMfieldofpointcharge}) in the expression for the angular momentum (\ref{angular momentum}), and dropping the self energy terms, one finds that the outgoing angular momentum is
\begin{align*}
&\lim_{c\rightarrow \infty }J_{\text{E.M.}}^{\beta\gamma}\Big(\set{u^{\mu}_{i}},n^{\mu},c\Big)=\sum_{\substack{i<j\\ i,j\,\in\, \text{out}}} \frac{e_ie_j}{16\pi^2}\int_{\Sigma_n}\DD^3\sigma \sqrt{|g_{\sigma\sigma}|} \theta(x^2)\\
 &\frac{{u_i}^{[\beta}x^{\gamma]}\Big((n\. x)(u_j\.x)-x^2(n\.u_j)\Big)+(i\leftrightarrow j)+n^{[\beta}x^{\gamma]}\Big(x^2(u_i\.u_j)-(u_i\.x)(u_j\.x)\Big)}{\Big((u_i\.x)^2-x^2\Big)^{\frac{3}{2}}\Big((u_j\.x)^2-x^2\Big)^{\frac{3}{2}}}.\numberthis\label{integral}
\end{align*}
In appendix \ref{app:evaluating integral} we evaluate (\ref{integral}) and find our main result,
\begin{equation}
  \boxed{\lim_{c\rightarrow \infty }J_{\text{E.M.}}^{\beta\gamma}\Big(\set{u^{\mu}_{i}},n^{\mu}\Big)=\sum_{\substack{i<j\\ i,j\,\in\, \text{out}}} \frac{e_ie_j}{4\pi}\frac{u^{[\beta}_iu^{\gamma]}_j}{\Big((u_i\.u_j)^2-1\Big)^{\frac{3}{2}}} \log\bigg(\frac{n\.u_i}{n\. u_j}\bigg).}\label{mainresult}
\end{equation}
In appendix \ref{app:instates} we discuss the angular momentum on a early time Cauchy surface $c \ll -1$ and find the same result as above but with an additional minus sign and we sum only over incoming charged particles
\begin{equation}
  \boxed{\lim_{c\rightarrow -\infty }J_{\text{E.M.}}^{\beta\gamma}\Big(\set{u^{\mu}_{i}},n^{\mu}\Big)=-\sum_{\substack{i<j\\ i,j\,\in\, \text{in}}} \frac{e_ie_j}{4\pi}\frac{u^{[\beta}_iu^{\gamma]}_j}{\Big((u_i\.u_j)^2-1\Big)^{\frac{3}{2}}} \log\bigg(\frac{n\.u_i}{n\. u_j}\bigg).}\numberthis\label{secondmainresult}
\end{equation}
The result (\ref{mainresult}-\ref{secondmainresult}) exhibits many interesting features. First notice that it is independent of $c$ which is our proxy for time. Hence there is a time-independent contribution to the EM angular momentum at asymptotic times, even when there are only charged particles in the asymptotic states. This observation suggests that the asymptotic states $\ket{\set{p_i^{\mu},\sigma_i,e_i}}$ used to describe the quantum scattering of charged particles should be further appended by states which account for the asymptotic angular momentum of the EM field. This is in line with the results of \cite{DiVecchia:2022owy} where it was demonstrated that the Faddeev-Kulish states carry a non-zero angular momentum. As the angular momentum tensor is promoted to the operator generating Lorentz transformations in the quantum theory, the observation that the angular momentum of the EM field approaches a non-zero value at asymptotic times suggests that the Lorentz transformation properties of the asymptotic states deviate from those of the free theory. For this reason it may be of interest to study the Lorentz transformation properties of the FK states. A striking feature of (\ref{mainresult}) is it's explicit dependence on the Cauchy surface via it's dependence on the normal vector $n^{\mu}$. Indeed, $J^{\beta\gamma}$ is only a tensor if one also transforms the Cauchy surface appropriately
\begin{equation}
J_{\text{E.M.}}^{\beta\gamma}\Big(\set{\Lambda^{\mu}_{\,\,\,\nu}u^{\nu}_{i}},\Lambda^{\mu}_{\,\,\,\nu}n^{\nu}\Big)=\Lambda^{\beta}_{\,\,\,\beta'}\Lambda^{\gamma}_{\,\,\,\gamma'}J_{\text{E.M.}}^{\beta'\gamma'}\Big(\set{u^{\mu}_{i}},n^{\mu}\Big).
\end{equation}
The dependence on $n^{\mu}$ is due to the fact that the mechanical and electromagnetic angular momentum densities are not independently conserved at asymptotic times,
\begin{equation}
  \d_{\alpha}M^{\alpha\beta\gamma}_{\text{E.M.}}=- \d_{\alpha}M^{\alpha\beta\gamma}_{\text{mech.}}=  -x^{[\gamma}F^{\beta] \lambda}J_{\lambda} ,
\end{equation}
as this torque scales logarithmically with the proper time of the source $\ln\tau$, indicating that even at asymptotic times there is a continual exchange of angular momentum between the particles and the EM field. Consequently it is only the sum of these two quantities which are conserved
\begin{equation}
\d_{\alpha}\Big(M^{\alpha\beta\gamma}_{\text{E.M.}}+M^{\alpha\beta\gamma}_{\text{mech.}}\Big)=0
\end{equation}
and therefore independent of the chosen Cauchy surface, which we verify in the next section. 
The relative sign between the in and out angular momenta has interesting implications. For large impact parameters where the particles remain undeflected and consequently have the same in-momenta as out-momenta, there still occurs a shift in the angular momentum of the electromagnetic field. This phenomenon was discovered in \cite{Gralla:2021eoi} where they they christened the effect as an ``electromagnetic scoot". The same authors found that gravitational systems also exhibit a ``gravitational scoot" \cite{Gralla:2021qaf}. It would be interesting to study whether these shifts are tied to any known memory effects. It is interesting to note that the angular momentum is independent of the possible length scales occurring in the problem. We hope to find a conformal symmetry argument for why (\ref{mainresult}-\ref{secondmainresult}), must be the result. We note that the angular momentum should be associated to the center-of-mass of the electromagnetic field. We are referring to the fact that for each pair of particles it is possible to go to a frame where their contribution to $J_{\text{E.M.}}^{ij}$'s rotational components vanish, namely the rest frame of one of the particles, whereas it is impossible to go to a frame where all three boost components $J^{0i}_{\text{E.M.}}$ vanish for a given pair of particles. This should be contrasted with the angular momentum present in the electromagnetic field in the presence of dyons (see (1.1) of \cite{Zwanziger:1972sx}) which is proportional to $J^{\alpha\beta}_{\text{E.M.}}=\e^{\alpha\beta}{}_{\gamma\delta}u_i^{\gamma}u_j^{\delta}$ and for visa-versa reasons is intrinsically associated to the rotational angular momentum of the electromagnetic field, and not it's center of mass. 
\section{Mechanical angular momentum}\label{sect:mech}
At asymptotic times the trajectories of charged particles deviate from linear motion by an order $\log(\tau)$ correction due to their electromagnetic interactions with the other charged particles,
\begin{gather}
\lim_{\tau_i\rightarrow \pm \infty }x_i^{\mu}=u^{\mu}_i\tau_i+b^{\mu}_i\mp\log|m\tau_i|\sum_{\substack{j \\ j\neq i}}\frac{e_ie_j}{4\pi}\frac{(u_i\.u_j)u_{i}^{\mu}-u^{\mu}_j}{m_i\Big((u_i\.u_j)^2-1\Big)^{\frac{3}{2}}}+\O\big(\tau^{-1}_i\log\tau_i\big)\label{motion}\\
\lim_{\tau_i\rightarrow \pm \infty }p_i^{\mu}=m_iu^{\mu}_i\mp\frac{1}{|\tau_i|}\sum_{\substack{j \\ j\neq i}}\frac{e_ie_j}{4\pi}\frac{(u_i\.u_j)u_{i}^{\mu}-u^{\mu}_j}{\Big((u_i\.u_j)^2-1\Big)^{\frac{3}{2}}}+\O\big(\tau^{-2}_i\log\tau_i\big)\label{momentum}
\end{gather}
see e.g. \cite{Sahoo:2018lxl,Zwanziger:1973if} for a derivation of this result. Here $m$ is an arbitrary constant which we choose to be the same for all particles, and will drop out of the final result. $b^{\mu}_i$ is an impact parameter. To compute the mechanical angular momentum of particle $i$,
\begin{equation}
J^{\beta\gamma}_{i,\text{mech.}}=p_i^{[\beta}x^{\gamma]}_{i}\label{mechangularmomentum}
\end{equation}
we need to solve for the $\tau_i$ at which the particle intersects the Cauchy surface $\Sigma_n$. Plugging in (\ref{motion}) into the defining equation for the hypersurface (\ref{cauchy surface}) and only retaining leading order terms in $c$ we find that the particle intersects the hypersurface at proper time
\begin{gather}
    n\.x_i=c\implies
    \tau_i=\frac{c}{n\.u_i}+\O\big(c^0\big).
\end{gather}
Plugging in this value for $\tau_i$ when using (\ref{motion}-\ref{momentum}) in (\ref{mechangularmomentum}), and then summing over all charged particles, we find that the total mechanical angular momentum is
\begin{equation}
    \lim_{c\rightarrow \pm \infty }  J^{\beta\gamma}_{\text{mech.}}\Big(\set{u_i},n\Big)=\sum_ip^{[\beta}_ib^{\gamma]}_i\mp\sum_{i<j}\frac{e_ie_j}{4\pi}\frac{u_i^{[\beta}u_j^{\gamma]}}{\Big((u_i\.u_j)^2-1\Big)^{\frac{3}{2}}}\log\Big(\frac{n\.u_i}{n\. u_j}\Big)\label{mech.ang.mom.}
\end{equation}
where the sum is only over in particles if $c<0$ and only over out particles if $c>0$. The first term is the familiar orbital angular momentum of a particle with momentum $p_i^{\mu}$ at impact parameter $b_i^{\mu}$. Note that the second term in (\ref{mech.ang.mom.}) precisely cancels the electromagnetic angular momentum (\ref{mainresult}). Indeed it must in order for the total angular momentum of the system to be conserved/independent of the Cauchy surface. 
\section*{Acknowledgements}
We thank Shounak De, Yangrui Hu, Lecheng Ren, Marcus Spradlin, Akshay Yelleshpur Srikant and Anastasia Volovich for helpful discussions. LL acknowledges support from the US Department of Energy contract DE–SC0010010 Task F.
\appendix
\section{Evaluating the angular momentum integral}\label{app:evaluating integral}
In this appendix we evaluate the integral (\ref{integral})
\begin{align*}
&I^{\beta\gamma}\Big(u^{\mu}_1,u^{\mu}_2,n^{\mu},c\Big)=\int_{\Sigma_n}\DD^3\sigma \sqrt{|g_{\sigma\sigma}|} \theta(x^2)\\
 &\frac{{u_1}^{[\beta}x^{\gamma]}\Big((n\. x)(u_2\.x)-x^2(n\.u_2)\Big)+(1\leftrightarrow 2)+n^{[\beta}x^{\gamma]}\Big(x^2(u_1\.u_2)-(u_1\.x)(u_2\.x)\Big)}{\Big((u_1\.x)^2-x^2\Big)^{\frac{3}{2}}\Big((u_2\.x)^2-x^2\Big)^{\frac{3}{2}}}.\numberthis\label{defofI}
\end{align*}
where for ease of demonstration we have selected the contributions from particles $i,j=1,2$ and factored out $\frac{e_1e_2}{16\pi^2}$. First we analyze the possible tensor structures for $I^{\beta\gamma}$ which is an anti-symmetric second rank tensor, and thus has potentially six independent components, which are formed from the vectors $n^{\mu},u_1^{\mu}$ and $u_2^{\mu}$. We will first show that the conformal properties of the integral restricts the possible tensor structures down to two. The $x^{\mu}$ in (\ref{defofI}) is understood to be evaluated in co-ordinates adapted to the hypersurface
\begin{gather}
    x^{\mu}=cn^{\mu}+\sum_{i=1}^3\sigma_i\e_i^{\mu}(n)\\
    \e_i\.n=0,\qquad \e_{i}\.\e_{j}=-\delta_{ij},\qquad \sqrt{|g_{\sigma\sigma}|}=1.
\end{gather}
The $\theta(x^2)=\theta(c^2-|\vec{\sigma}|^2)$ then restricts the integration range of the $\sigma$'s to the solid ball of radius $c$. Changing to integration co-ordinates $\sigma_i'=\frac{\sigma_i}{c}$ we see that the six powers of $c$ in the denominator of (\ref{defofI}) precisely cancel the six powers of $c$ in the numerator of (\ref{defofI}). This proves that (\ref{defofI}) is independent of $c$, and consequently via the defining equation of the hypersurface $n\.x=c$ that (\ref{defofI}) can only depend on ratio's of $n^{\mu}$. Secondly, let us for a moment generalize the definition of (\ref{defofI}) by making the replacement in the denominators
\begin{equation}
    \Big((u_i\.x)^2-x^2\Big)^{-\frac{3}{2}}\rightarrow \Big((u_i\.x)^2-u_i^2x^2\Big)^{-\frac{3}{2}}
\end{equation}
for $i=1,2$ and we no longer restrict $u_i^2=1$, we only require that $u_i^{\mu}$ remain timelike. We leave the numerators untouched. This defines a new integral $\tilde{I}^{\beta\gamma}$ from which we can readily obtain our desired integral
\begin{equation}
I^{\beta\gamma}\Big(u^{\mu}_1,u^{\mu}_2,n^{\mu}\Big)=\tilde{I}^{\beta\gamma}\Big(u^{\mu}_1,u^{\mu}_2,n^{\mu}\Big)\at{u_1^2=u_2^2=1}.
\end{equation}
From this equation we can trivially obtain the tensor structure of $I$ from that of $\tilde{I}$. Now note that $\tilde{I}$ is homogeneously of degree $-2$ in each of the four velocities
\begin{gather}
    \tilde{I}(\alpha u_1^{\mu},u_2^{\mu},n^{\mu})=\alpha^{-2}\tilde{I}(u_1^{\mu},u_2^{\mu},n^{\mu}),\qquad     \tilde{I}( u_1^{\mu},\alpha u_2^{\mu},n^{\mu})=\alpha^{-2}\tilde{I}(u_1^{\mu},u_2^{\mu},n^{\mu})
\end{gather}
for $\alpha>0$ a positive real number. Only two tensor structures are compatible with all of the constraints we have listed so far,
\begin{equation}
    I^{\beta\gamma}(u_1^{\mu},u_2^{\mu},n^{\mu})=f_1u^{[\beta}_{1}u_2^{\gamma]}+f_2\e^{\beta\gamma}{}_{\delta\lambda}u^{\delta}_1u^{\lambda}_2\label{tensordecomp}
\end{equation}
where $f_i\big(u_1\.u_2,n\.u_1,n\.u_2\big)$ are Lorentz invariant functions. Note that (\ref{defofI}) is symmetric under $(1\leftrightarrow 2)$ and consequently we have that the $f_i$'s must be anti-symmetric under $(1\leftrightarrow 2)$. We can project onto the individual tensor components in (\ref{tensordecomp}) by contracting with the appropriate tensors,
\begin{align}
    f_1=\frac{u^{\beta}_1u^{\gamma}_2I_{\beta\gamma}}{1-(u_1\.u_2)^2},\qquad f_2=-\frac{1}{2}\frac{u_1^{\delta}u_2^{\lambda}\e_{\delta\lambda\beta\gamma}I^{\beta\gamma}}{1-(u_1\.u_2)^2}.
\end{align}
Explicit evaluation of the integral reveals that $f_2=0$. We now evaluate $f_1$. Contracting $u^{\beta}_1u^{\gamma}_2$ onto $I_{\beta\gamma}$ in (\ref{defofI}) and performing some algebra to manipulate the numerator into a convenient form we find
\begin{align}
f_1&=\frac{1}{1-(u_1\.u_2)^2}\int_{\Sigma_n}\DD^3\sigma \sqrt{|g_{\sigma\sigma}|} \theta(x^2)\frac{\Big((u_1\.x)^2-x^2\Big)\Big((n\.u_2)(u_2\.x)-(n\.x)\Big)-(1\leftrightarrow 2)}{\Big((u_1\.x)^2-x^2\Big)^{\frac{3}{2}}\Big((u_2\.x)^2-x^2\Big)^{\frac{3}{2}}}\\
&=\frac{1}{1-(u_1\.u_2)^2}\int_{\Sigma_n}\DD^3\sigma \sqrt{|g_{\sigma\sigma}|} \theta(x^2)\frac{{u^2}_{\mu}n_{\nu}x^{[\mu}{u_2}^{\nu]}}{\Big((u_1\.x)^2-x^2\Big)^{\frac{1}{2}}\Big((u_2\.x)^2-x^2\Big)^{\frac{3}{2}}}-(1\leftrightarrow 2).\label{f1}
\end{align}
In the second equality we have cancelled repeated powers of the same function in the numerator and denominator as well as manipulated the numerators into a form that reveals that the integrand contains terms proportional to the field strengths of the particles (\ref{EMfieldofpointcharge}). Defining the scaled field strengths,
\begin{equation}
    G^{\mu\nu}_i=\frac{x^{[\mu}{u_i}^{\nu]}}{\Big((u_i\.x)^2-x^2\Big)^{\frac{3}{2}}}
\end{equation}
which due to Maxwell's equations $\d_{\mu}F^{\mu\nu}=J^{\nu}$ satisfies,
\begin{equation}
    \d_{\nu}G^{\mu\nu}_i=-4\pi u^{\nu}_2 \int_0^{\infty}\DD\tau\,\, \delta^4(x^{\alpha}-u_2^{\alpha}\tau)\label{divofG}.
\end{equation}
We then have that
\begin{gather}
    f_1=\frac{1}{1-(u_1\.u_2)^2}\int_{\Sigma_n}\DD^3\sigma \sqrt{|g_{\sigma\sigma}|} \theta(x^2)\frac{{u^2}_{\mu}n_{\nu}G^{\mu\nu}_2}{\Big((u_1\.x)^2-x^2\Big)^{\frac{1}{2}}}-(1\leftrightarrow 2).\label{intermediate}
\end{gather}
The divergence property (\ref{divofG}) motivates us to use Stokes theorem and interpret $\Sigma_n$ as part of a boundary enclosing a $4$-volume. To this end we need to add extra boundary terms. We would like the added boundary terms to evaluate to zero. To this end let us note $f_1$ vanishes on the surface which has a normal vector
\begin{equation}
    {n'}^{\mu}=\frac{1}{\sqrt{2}}\frac{u_1^{\mu}+u_2^{\mu}}{\sqrt{(u_1\.u_2)^2+1}}
\end{equation}
which is easily deduced from the fact that $f_1(u_1\.n,u_2\.n,n)$ is anti-symmetric under $1\leftrightarrow 2$. We therefore add the boundary $\Sigma_{n'}$ below\footnote{We add the surface $\Sigma_{n'}$ below $\Sigma_{n}$ so that the normal vector $n^{\mu}$ does not flip sign when using Stokes theorem} $\Sigma_n$, i.e. $c'<c$, and join the two boundaries at radial infinity, which we are free to do without affecting the integral $f_1$ as the integrand (\ref{intermediate}) falls off sufficiently fast $\frac{1}{R^3}$ at radial infinity. We call the $4-$volume bounded by this closed surface $V_{\Sigma}$. Using Stokes theorem we have that
\begin{align}
    f_1&=\frac{1}{1-(u_1\.u_2)^2}\int_{V_{\Sigma}}\DD^4x \d_{\nu}\Bigg(\theta(x^2)\frac{{u^2}_{\mu}G^{\mu\nu}_2}{\Big((u_1\.x)^2-x^2\Big)^{\frac{1}{2}}}\Bigg)-(1\leftrightarrow 2)\\
    &=\frac{1}{1-(u_1\.u_2)^2}\int_{V_{\Sigma}}\DD^4x \Bigg(\theta(x^2)\frac{{u^2}_{\mu}\d_{\nu}G^{\mu\nu}_2}{\Big((u_1\.x)^2-x^2\Big)^{\frac{1}{2}}}\Bigg)-(1\leftrightarrow 2)\\
    &=\frac{4\pi}{(u_1\.u_2)^2-1}\int_{V_{\Sigma}}\DD^4x\int_0^{\infty}\DD\tau_2\Bigg(\theta(x^2)\frac{\delta^4(x^{\alpha}-u_2^{\alpha}\tau_2) }{\Big((u_1\.x)^2-x^2\Big)^{\frac{1}{2}}}\Bigg)-(1\leftrightarrow 2)\label{bounds}\\
    &=\frac{4\pi}{\Big((u_1\.u_2)^2-1\Big)^{\frac{3}{2}}}\int^{u^{\alpha}_{2}\tau_2 \cap \Sigma_{n}}_{u^{\alpha}_{2}\tau_2 \cap \Sigma_{n'}}\frac{\DD\tau}{|\tau|}-(1\leftrightarrow 2)\label{intersecteqns}
\end{align}
In going to the second equality we used that the terms where the derivative does not act on $G^{\mu\nu}_2$ are symmetric under $(1\leftrightarrow 2)$ and hence vanish when we antisymmetrize. In the third equality we have used the divergence property (\ref{divofG}). At the fourth equality the bounds on the $\tau_2$ integral indicate that we integrate $\tau$ from the proper time at which particle $2$ intersects the lower boundary, until the proper time that it intersects the upper boundary. To find the upper and lower intersection times one simply uses the defining equations for the boundaries, e.g. $(n\.x)=c$, to find 
\begin{gather}
    \tau^{\text{lower}}_2=\frac{c'}{n'\. u_2},\qquad \tau^{\text{upper}}_2=\frac{c}{n\.u_2}.
\end{gather}
Note that ${n'}\.u_2$ is symmetric in $(1\leftrightarrow 2)$. Altogether then we have that\footnote{The sign function arises from the relation $\int_A^B\frac{\DD}{|x|}=\text{sign}(B)\ln|\frac{B}{A}|$ where the interval $[A,B]$ does not contain $0$ so that no poles are encountered along the integration contour.}
\begin{equation}
    f_1=\text{sign}(c)\frac{4\pi}{\Big((u_1\.u_2)^2-1\Big)^{\frac{3}{2}}}\log\Big(\frac{n\.u_1}{n\.u_2}\Big)
\end{equation}
Then using the relation between $f_1$ and our integral $I^{\beta\gamma}$ (\ref{tensordecomp}) we have that
\begin{equation}
I^{\beta\gamma}\Big(u_1^{\mu},u_2^{\mu},n^{\mu}\Big)=4\pi\text{sign}(c)\,\,\frac{u^{[\beta}_{1}u^{\gamma]}_{2}}{\Big((u_1\.u_2)^2-1\Big)^{\frac{3}{2}}}\log\Big(\frac{n\.u_1}{n\.u_2}\Big).\label{Ifinal}
\end{equation}
Finally, to obtain the angular momentum we multiply by $\frac{e_1e_2}{16\pi^2}$ and sum over all pairs of outgoing particles to obtain our main result (\ref{mainresult}).
\addtocontents{toc}{\protect\setcounter{tocdepth}{1}}%
\section{Contributions from the in-states}\label{app:instates}
\subsection{In-states do not contribute to the out angular momentum}
The field strength's produced by incoming currents have support on the outgoing Cauchy surface (outgoing Cauchy surface is defined as a $c>0$) which could thus potentially affect the angular momentum on an outgoing surface. However we will show here that these terms give zero contribution.
We model an in-current by simply changing the bounds of integration from that of an out-current (\ref{currents})
\begin{equation}
    J^{\mu}_{i}=e_iu^{\mu}_{i}\int_{-\infty}^{0}\DD \tau \,\,\delta^{4}\Big(x^{\mu}-u^{\mu}_{i}\tau-b_i^{\mu}+\O(\log\tau)\Big),\qquad u^{\mu}_{i}=\frac{p^{\mu}_{i}}{m_{i}}.\label{incurrents}
\end{equation}
The field strength created by such an in-source has support everywhere except for the forward lightcone of the origin (using the retarded Green's function)
\begin{equation}
  \lim_{|t|\rightarrow \infty}  F^{\mu\nu}_{i}=\frac{e_i}{4\pi}\frac{x_{}^{{[}\mu}{u_i}^{\nu{]}}}{\Big((u_i\. x)^2-x^2\Big)^{\frac{3}{2}}}(1-\theta(x^0)\theta(x^2))+\O(t^{-3}\log(t)).\label{infield}
\end{equation}
On an outgoing Cauchy surface $c>0$ the contributions of in-currents to the angular momentum on such an outgoing Cauchy surface can be obtained simply by making the replacement
$\theta(x^2)\rightarrow \theta(-x^2)$ in (\ref{integral}). We thus seek to show that (\ref{defofI}) is zero when making this replacement. Following the reasoning of appendix \ref{app:evaluating integral} we see that the arguments which lead from (\ref{defofI}) to (\ref{intersecteqns}) were independent of the $\theta(x^2)$. We are thus lead to the conclusion that the in-currents contributions to the out-going angular momentum is
\begin{align}
    f_1&=\frac{4\pi}{(u_1\.u_2)^2-1}\int_{V_{\Sigma}}\DD^4x\int_{-\infty}^0\DD\tau_2\Bigg(\theta(-x^2)\frac{\delta^4(x^{\alpha}-u_2^{\alpha}\tau_2) }{\Big((u_1\.x)^2-x^2\Big)^{\frac{1}{2}}}\Bigg)-(1\leftrightarrow 2)\\
    &=0
\end{align}
which is zero because $\theta(-\tau^2)=0$ or more simply put, because the in-current has no support within the 4-volume $V_{\Sigma}$. We therefore conclude that the incoming currents do not contribute to the outgoing angular momentum.
\subsection{In-states contribution to the in angular momentum}
Here we compute the angular momentum on an early time Cauchy surface $c\ll 0$ due to the incoming currents (\ref{incurrents}). The field strengths created by the in-sources (\ref{infield}) have support everywhere on the $c\ll 0$ Cauchy surface. Consequently the integral expression for the electromagnetic angular momentum on an in-Cauchy surface due to the in-currents is the same as in (\ref{integral}), except we drop the $\theta(x^2)$, sum over only incoming particles, and $\Sigma_{n}$ must now be taken to be the in-Cauchy surface. We therefore seek to evaluate (\ref{defofI}) with these replacements. The arguments that lead from (\ref{defofI}) to (\ref{Ifinal}) can be repeated verbatim with these replacements. The only difference at an intermediate step are the bounds of integration for the $\tau$ integral at (\ref{bounds}) would be $[-\infty,0]$ but this is inconsequential. We conclude that the in particles contribution to the incoming momentum is (\ref{secondmainresult}).
\bibliography{asym.bib}

\providecommand{\href}[2]{#2}\begingroup\raggedright\begin{thebibliography}{10}

\bibitem{dollard}
J.~D. Dollard, {\it Asymptotic convergence and the coulomb interaction},  {\em
  Journal of Mathematical Physics} {\bf 5} (1964), no.~6 729--738,
  [\href{http://arxiv.org/abs/https://doi.org/10.1063/1.1704171}{{\tt
  https://doi.org/10.1063/1.1704171}}].

\bibitem{Chung:1965zza}
V.~Chung, {\it {Infrared Divergence in Quantum Electrodynamics}},  {\em Phys.
  Rev.} {\bf 140} (1965) B1110--B1122.

\bibitem{Kibble:1968sfb}
T.~W.~B. Kibble, {\it {Coherent Soft-Photon States and Infrared Divergences. I.
  Classical Currents}},  {\em J. Math. Phys.} {\bf 9} (1968), no.~2 315--324.

\bibitem{Kulish:1970ut}
P.~P. Kulish and L.~D. Faddeev, {\it {Asymptotic conditions and infrared
  divergences in quantum electrodynamics}},  {\em Theor. Math. Phys.} {\bf 4}
  (1970) 745.

\bibitem{Caron-Huot:2022jli}
S.~Caron-Huot, Y.-Z. Li, J.~Parra-Martinez, and D.~Simmons-Duffin, {\it
  {Graviton partial waves and causality in higher dimensions}},
  \href{http://arxiv.org/abs/2205.01495}{{\tt arXiv:2205.01495}}.

\bibitem{Caron-Huot:2022ugt}
S.~Caron-Huot, Y.-Z. Li, J.~Parra-Martinez, and D.~Simmons-Duffin, {\it
  {Causality constraints on corrections to Einstein gravity}},
  \href{http://arxiv.org/abs/2201.06602}{{\tt arXiv:2201.06602}}.

\bibitem{DiVecchia:2022owy}
P.~Di~Vecchia, C.~Heissenberg, and R.~Russo, {\it {Angular momentum of
  zero-frequency gravitons}},  \href{http://arxiv.org/abs/2203.11915}{{\tt
  arXiv:2203.11915}}.

\bibitem{Low:1954kd}
F.~E. Low, {\it {Scattering of light of very low frequency by systems of spin
  1/2}},  {\em Phys. Rev.} {\bf 96} (1954) 1428--1432.

\bibitem{Low:1958sn}
F.~E. Low, {\it {Bremsstrahlung of very low-energy quanta in elementary
  particle collisions}},  {\em Phys. Rev.} {\bf 110} (1958) 974--977.

\bibitem{Burnett:1967km}
T.~H. Burnett and N.~M. Kroll, {\it {Extension of the low soft photon
  theorem}},  {\em Phys. Rev. Lett.} {\bf 20} (1968) 86.

\bibitem{Gell-Mann:1954wra}
M.~Gell-Mann and M.~L. Goldberger, {\it {Scattering of low-energy photons by
  particles of spin 1/2}},  {\em Phys. Rev.} {\bf 96} (1954) 1433--1438.

\bibitem{Lysov:2014csa}
V.~Lysov, S.~Pasterski, and A.~Strominger, {\it {Low\textquoteright{}s
  Subleading Soft Theorem as a Symmetry of QED}},  {\em Phys. Rev. Lett.} {\bf
  113} (2014), no.~11 111601, [\href{http://arxiv.org/abs/1407.3814}{{\tt
  arXiv:1407.3814}}].

\bibitem{Sahoo:2018lxl}
B.~Sahoo and A.~Sen, {\it {Classical and Quantum Results on Logarithmic Terms
  in the Soft Theorem in Four Dimensions}},  {\em JHEP} {\bf 02} (2019) 086,
  [\href{http://arxiv.org/abs/1808.03288}{{\tt arXiv:1808.03288}}].

\bibitem{Bieri:2013hqa}
L.~Bieri and D.~Garfinkle, {\it {An electromagnetic analogue of gravitational
  wave memory}},  {\em Class. Quant. Grav.} {\bf 30} (2013) 195009,
  [\href{http://arxiv.org/abs/1307.5098}{{\tt arXiv:1307.5098}}].

\bibitem{Pasterski:2015zua}
S.~Pasterski, {\it {Asymptotic Symmetries and Electromagnetic Memory}},  {\em
  JHEP} {\bf 09} (2017) 154, [\href{http://arxiv.org/abs/1505.00716}{{\tt
  arXiv:1505.00716}}].

\bibitem{AtulBhatkar:2020hqz}
S.~Atul~Bhatkar, {\it {New asymptotic conservation laws forelectromagnetism}},
  {\em JHEP} {\bf 02} (2021) 082, [\href{http://arxiv.org/abs/2007.03627}{{\tt
  arXiv:2007.03627}}].

\bibitem{Hirai:2018ijc}
H.~Hirai and S.~Sugishita, {\it {Conservation Laws from Asymptotic Symmetry and
  Subleading Charges in QED}},  {\em JHEP} {\bf 07} (2018) 122,
  [\href{http://arxiv.org/abs/1805.05651}{{\tt arXiv:1805.05651}}].

\bibitem{Campiglia:2016hvg}
M.~Campiglia and A.~Laddha, {\it {Subleading soft photons and large gauge
  transformations}},  {\em JHEP} {\bf 11} (2016) 012,
  [\href{http://arxiv.org/abs/1605.09677}{{\tt arXiv:1605.09677}}].

\bibitem{Himwich}
E.~Himwich and A.~Strominger, {\it Celestial current algebra from low’s
  subleading soft theorem},  {\em Physical Review D} {\bf 100} (9, 2019).

\bibitem{Campiglia:2019wxe}
M.~Campiglia and A.~Laddha, {\it {Loop Corrected Soft Photon Theorem as a Ward
  Identity}},  {\em JHEP} {\bf 10} (2019) 287,
  [\href{http://arxiv.org/abs/1903.09133}{{\tt arXiv:1903.09133}}].

\bibitem{Donnay:2022sdg}
L.~Donnay, S.~Pasterski, and A.~Puhm, {\it {Goldilocks modes and the three
  scattering bases}},  {\em JHEP} {\bf 06} (2022) 124,
  [\href{http://arxiv.org/abs/2202.11127}{{\tt arXiv:2202.11127}}].

\bibitem{Gralla:2021eoi}
S.~E. Gralla and K.~Lobo, {\it {Electromagnetic scoot}},  {\em Phys. Rev. D}
  {\bf 105} (2022), no.~8 084053, [\href{http://arxiv.org/abs/2112.01729}{{\tt
  arXiv:2112.01729}}].

\bibitem{Zwanziger:1972sx}
D.~Zwanziger, {\it {Angular distributions and a selection rule in charge-pole
  reactions}},  {\em Phys. Rev. D} {\bf 6} (1972) 458--470.

\bibitem{Csaki1}
C.~Csaki, S.~Hong, Y.~Shirman, O.~Telem, J.~Terning, and M.~Waterbury, {\it
  {Scattering amplitudes for monopoles: pairwise little group and pairwise
  helicity}},  {\em JHEP} {\bf 08} (2021) 029,
  [\href{http://arxiv.org/abs/2009.14213}{{\tt arXiv:2009.14213}}].

\bibitem{Csaki2}
C.~Cs\'aki, S.~Hong, Y.~Shirman, O.~Telem, and J.~Terning, {\it {Completing
  Multiparticle Representations of the Poincar\'e Group}},  {\em Phys. Rev.
  Lett.} {\bf 127} (2021), no.~4 041601,
  [\href{http://arxiv.org/abs/2010.13794}{{\tt arXiv:2010.13794}}].

\bibitem{Lippstreu:2021avq}
L.~Lippstreu, {\it {Zwanziger\textquoteright{}s pairwise little group on the
  celestial sphere}},  {\em JHEP} {\bf 11} (2021) 051,
  [\href{http://arxiv.org/abs/2106.00084}{{\tt arXiv:2106.00084}}].

\bibitem{Gralla:2021qaf}
S.~E. Gralla and K.~Lobo, {\it {Self-force effects in post-Minkowskian
  scattering}},  {\em Class. Quant. Grav.} {\bf 39} (2022), no.~9 095001,
  [\href{http://arxiv.org/abs/2110.08681}{{\tt arXiv:2110.08681}}].

\bibitem{Zwanziger:1973if}
D.~Zwanziger, {\it {Reduction formulas for charged particles and coherent
  states in quantum electrodynamics}},  {\em Phys. Rev. D} {\bf 7} (1973)
  1082--1099.

\end{thebibliography}\endgroup
\bibliographystyle{JHEP}
\end{document}